\documentclass[physrev,reprint,bibnotes,citeautoscript,nobibnotes]{revtex4-2}  

\usepackage[T1]{fontenc}
\usepackage[utf8]{inputenc}
\usepackage[australian]{babel}
\usepackage[a4paper,centering,hmargin=1.75cm,vmargin=2cm]{geometry} 
\usepackage{amsmath,amssymb,graphicx,bm,microtype}
\usepackage[dvipsnames]{xcolor}
\usepackage[colorlinks,allcolors=blue!50!black]{hyperref}
\usepackage[all]{hypcap}
\usepackage{cleveref}
\usepackage{braket}
\usepackage{siunitx}

\begin{document}

\title{Fast and Accurate Simulations of Partially Delocalised Charge Separation in Organic Semiconductors}

\author{Jacob T. Willson}
\author{Daniel Balzer}
\author{Ivan Kassal*}
\affiliation{School of Chemistry, University of Sydney, NSW 2006, Australia\\
*\;E-mail: ivan.kassal@sydney.edu.au}

\begin{abstract}
\begin{center}\textbf{Abstract}\end{center}
Accurate computational screening of candidate materials promises to accelerate the discovery of higher-efficiency organic photovoltaics (OPVs). However, modelling charge separation in OPVs is challenging because accurate models must include disorder, polaron formation, and charge delocalisation. Delocalised kinetic Monte Carlo (dKMC) includes these three essential ingredients, but it suffers from high computational cost. Recently, we developed jumping kinetic Monte Carlo (jKMC), a computationally cheap and accurate model of delocalised charge transport that models transport over a lattice of identical, spherical polarons. Here, we extend jKMC to describe the separation of a charge-transfer state, showing that this simplified approach can reproduce the considerable improvements in charge-separation efficiencies caused by delocalisation and first seen in dKMC. The low computational cost and simplicity of jKMC allows it to be applied to parameter regimes intractable by dKMC, and ensures jKMC can be easily incorporated into any existing KMC model. 
\begin{center}
\includegraphics[width=5cm]{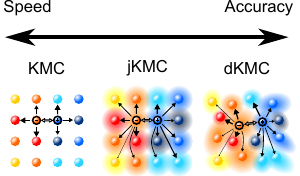}
\end{center}
\end{abstract}

\maketitle

Organic photovoltaics (OPVs) benefit from the chemical tunability of organic molecules and polymers, which allow them to be optimised for specific applications~\cite{Kohler&Bassler}. However, exploring the large chemical space to discover new OPV materials is a monumental task, especially when coupled with the experimental effort required to construct and test OPV devices. This task can be assisted by fast and accurate computational screening to help identify new candidate materials~\cite{Zojer2021}.

A critical performance measure for any OPV is the separation efficiency of electron-hole pairs. OPVs are typically blends of electron-donor and -acceptor materials self-assembled into domains~\cite{Kohler&Bassler,Clarke2010}. Light absorption creates an exciton that diffuses towards the donor-acceptor interface, where it dissociates to form a Coulombically bound charge-transfer (CT) state. The separation of CT states into free charges can occur with near-unity internal quantum efficiency (IQE)~\cite{Park2009}, despite the available thermal energy being an order of magnitude smaller than the Coulomb binding energy~\cite{Nelson2011,Few2015}.

Charge delocalisation is widely thought to enhance the IQE of charge separation, a view supported by experimental observations of fast and efficient separation through delocalised CT states~\cite{Bakulin2012,Jailaubekov2013,Grancini2013,Gelinas2014,Falke2014,Tamai2017}. Delocalisation in OPVs is partial: disorder~\cite{Anderson1958} and polaron formation~\cite{Grover1971,Rice2018} localise charges, and partially delocalised states arise when this localisation is insufficient to reduce the state to one molecule. The open question is whether and how the typically small amount of delocalisation can improve device performance.

However, delocalisation considerably complicates the computational task of predicting charge-separation efficiencies, which requires modelling the dynamics of Coulombically bound electrons and holes as they separate. The partial delocalisation in OPVs means that charge dynamics typically occurs in the intermediate regime between the theoretically well-understood limits of delocalised band conduction and localised hopping~\cite{Kohler&Bassler,Oberhofer2017,Balzer2021,Balzer2022}. Including delocalisation in OPV modelling is essential because conventional localised-hopping models struggle to explain the near-unity IQEs observed experimentally.

The most complete charge-separation models for OPVs are atomistic simulations~\cite{Tamura2013,Huix2015,Polkehn2018,Peng2022}, used to model quantum-mechanical dynamics on small systems and short timescales. These detailed, ab initio models explain the ultrafast dissociation of electron-hole pairs, but cannot reach the mesoscopic scales relevant to device performance due to the large computational cost of tracking wavefunctions in an exponentially growing Hilbert space. For longer timescales, effective-Hamiltonian models of charge separation~\cite{Bittner2014,Jankovic2020,Balzer2022} can simulate larger systems by tracking fewer parameters, but still suffer from an exponentially growing Hilbert space, restricting them to describing short-time dynamics in one or two dimensions. These limitations make these models difficult to use in mesoscopic parameter scans required for computational screening.

\begin{figure*}
    \centering
    \includegraphics[width=\textwidth]{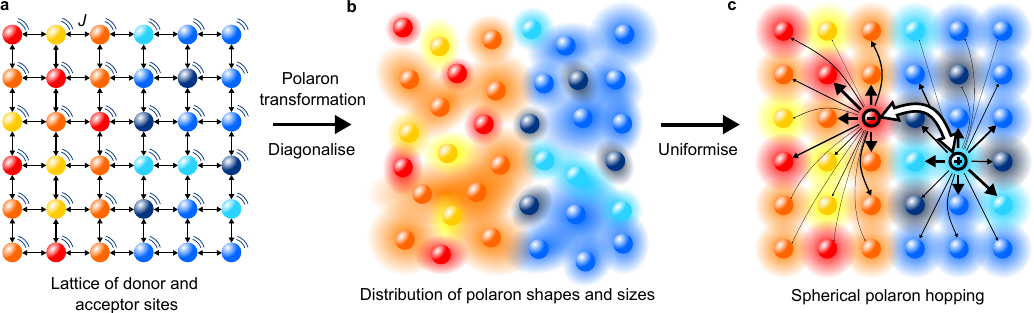}
    \caption{\textbf{The jKMC model of partially delocalised CT-state separation in disordered materials.} \textbf{(a)} The starting point of jKMC is a lattice of sites with disordered energies (different shades of colours), nearest-neighbour couplings $J$, and coupled to the environment (motion lines). A planar heterojunction is modelled by restricting the electron and the hole to the donor (blue) and acceptor (orange) lattice sites, respectively. \textbf{(b)} Diagonalising the Hamiltonian (in the polaron frame, and over the donor and acceptor domains separately) yields partially delocalised polarons with a distribution of shapes and sizes. \textbf{(c)} jKMC uniformises the polarons to model transport over a lattice of identical, spherically delocalised states. This delocalisation allows polarons to jump over their nearest neighbours and find more paths towards charge separation. Delocalisation also facilitates long-range recombination of the polarons (white arrow).}
    \label{fig:CS_jKMC_model}
\end{figure*}

Delocalised kinetic Monte Carlo (dKMC) is an effective-Hamiltonian model that was the first to include delocalisation, disorder, and polaron formation on a mesoscopic scale in three dimensions, demonstrating that delocalisation significantly enhances charge separation~\cite{Balzer2021,Balzer2022}. dKMC also explains how this enhancement occurs. Previously, the commonly assumed mechanism was that delocalisation increases the initial separation of the electron and hole, thereby reducing the Coulombic attraction between them~\cite{Few2015}. Instead of this energetic effect, dKMC shows that delocalisation enhances charge separation through a kinetic effect of faster charge transport~\cite{Balzer2022}. However, dKMC is still computationally expensive and limited to small amounts of delocalisation, meaning that wider use of delocalisation models requires a simpler model that captures the same kinetic effects.

Recently, we developed jumping kinetic Monte Carlo (jKMC)~\cite{Willson2023}, a model of partially delocalised charge transport that strikes a balance between computational speed and accuracy while allowing delocalisation, disorder, and polarons to be treated on mesoscopic scales. For single-carrier transport, it has nearly the same computational cost as conventional kinetic Monte Carlo (KMC), but an accuracy approaching that of dKMC. jKMC achieves its performance by modelling charge transport over a lattice of identical, spherically delocalised polarons to yield an easy-to-calculate hopping rate. The delocalisation allows polarons to jump beyond nearest neighbours, with a rate that is the Marcus rate~\cite{Marcus1956} modified with a distance-dependent exponential prefactor, $\frac{d}{a}e^{-2(d-a)/r_\mathrm{deloc}}$. This correction is similar to the phenomenological factor $e^{-2\gamma d}$  often used in Miller-Abrahams hopping rates~\cite{Miller1960}, except that it is rigorously justified, as opposed to being a fitting parameter~\cite{Bassler1993,Deibel2009,Schwarz2013,Nenashev2011,Tscheuschner2015,Athanasopoulos2017,Athanasopoulos2019}. 

Here, we extend jKMC to the two-body problem of charge separation and show that it reproduces the separation efficiencies of dKMC, significantly higher than the corresponding KMC models. With its low computational cost, jKMC can be applied in parameter regimes inaccessible to dKMC and the two-body version retains the similarity to KMC that means it can be used in any KMC code.

\section{Methods}

For simplicity, we model an OPV as a planar donor-acceptor heterojunction organised as a lattice of sites (\cref{fig:CS_jKMC_model}a). Disorder is introduced through assigning each site an independent random energy from a Gaussian density of states (DOS) of width $\sigma$~\cite{Bassler1993}. We assume nearest-neighbour coupling for the lattice, and that each site is linearly coupled to an identical, independent bath of harmonic oscillators. The donor and acceptor are set to have identical material parameters except that their DOSs are assumed to be sufficiently separated that the hole and electron are restricted to their respective domains. This assumption allows the charges to only interact via the Coulomb attraction or electron-hole recombination to the ground state. 

The full Hamiltonian of the model is unimportant for what follows, but is given in our previous dKMC work~\cite{Balzer2022}. Its only important feature is that, because of the disorder, its eigenstates have irregular shapes, off-lattice positions, and varying amounts of delocalisation (\cref{fig:CS_jKMC_model}b), all of which make modelling transport difficult. In dKMC, the eigenstates are found by diagonalising the system Hamiltonian after applying the polaron transformation~\cite{Balzer2021,Balzer2022}. 

In jKMC, to reduce the computational cost, we uniformise the polaron eigenstates by assuming that they are identical, exponentially decaying, spherical polarons whose centres are located on the original cubic lattice (\cref{fig:CS_jKMC_model}c)~\cite{Willson2023}. Polaron state $\ket{\nu}$  is described by the wavefunction
\begin{equation}
    \label{eqn:Spherical_polaron_approximation}
    \ket{\nu}=A \sum_i  \exp\left(-\frac{d_{i \nu}}{r_\mathrm{deloc}}\right)\ket{i},
\end{equation}
where the delocalisation radius $r_\mathrm{deloc}$ describes the extent of delocalisation, the sum runs over all lattice sites $i$, $d_{i\nu}$ is the distance between site $i$ and the centre of the polaron, and $A=\left(\sum_i \exp\left(-2d_{i \nu}/r_\mathrm{deloc}\right)\right)^{-1/2}$ is the normalisation. The delocalisation radius $r_\mathrm{deloc}$ must be chosen to capture the effective delocalisation in the system, based on a sample of polaron eigenstates~\cite{Balzer2021,Balzer2022}.

In the two-particle problem of charge separation, the Hamiltonian describes the joint Hilbert space of the electron and the hole, and its eigenstates (in the polaron frame) are joint electron-hole polaron states~\cite{Balzer2022}. Finding them by diagonalisation, as is done in dKMC~\cite{Balzer2022}, is an expensive calculation that, in three dimensions, is only possible in small systems. To reduce this cost, we assume in jKMC that each joint electron-hole state is separable, $\ket{\nu}=\ket{\nu_D,\nu_A}$, a product of independent polaron states $\ket{\nu_D}$ of the hole and $\ket{\nu_A}$ of the electron, which can be obtained by diagonalising the donor and acceptor Hamiltonians separately (\cref{fig:CS_jKMC_model}b). This approximation is assisted by the restriction of the electron and hole to separate domains, and the low likelihood that entanglement between them would persist on the long time scales of interest in jKMC.

To begin the simulation, we initialise the charges in an interfacial CT state, i.e., in electron and hole polarons centred on a nearest-neighbour pair of sites lying across the donor-acceptor interface, with the hole in the donor and and the electron in the acceptor. Following Balzer and Kassal~\cite{Balzer2022}, we compare two different methods of selecting the initial interfacial CT state: the random-initialisation method chooses one CT state uniformly at random, while the thermalised-initialisation method selects the state out of the thermally weighted Boltzmann distribution.

\begin{figure}
    \centering
    \includegraphics[width=\columnwidth]{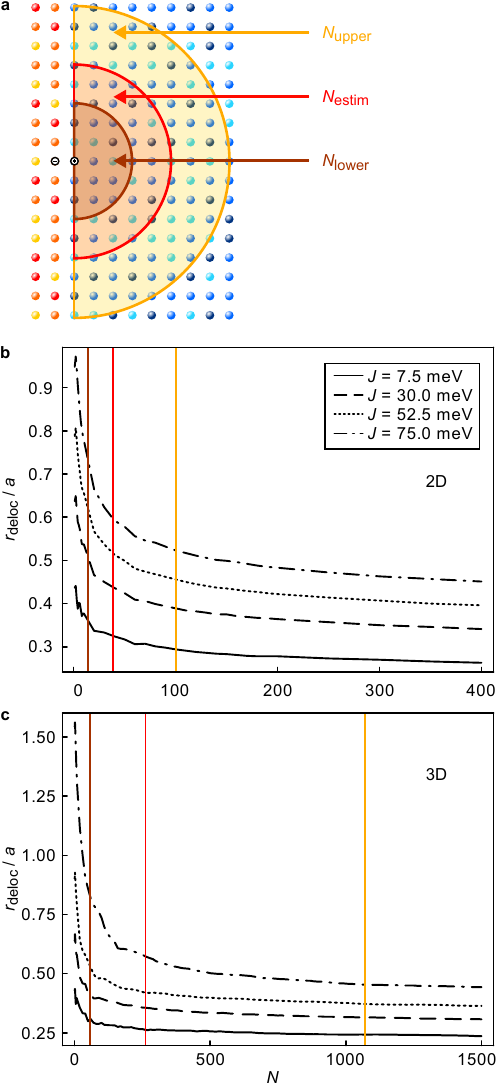}
    \caption{\textbf{Estimating the delocalisation radius $r_\mathrm{deloc}$ from the neighbourhood size $N$.} \textbf{(a)} $r_\mathrm{deloc}$ depends on the number of sites accessible to either charge during the charge separation, because larger $N$ increases the chance of finding localised energetic traps in the disordered DOS. We set $N=N_\mathrm{estim}$ equal to the number of sites inside a hemisphere of radius $r_N$, taken to be equal to the charge separation cutoff, which we assume to be $r_\mathrm{sep}=\SI{5}{nm}$. Due to the uncertainty of our estimate $N_\mathrm{estim}$, we give a range from $N_\mathrm{lower}$ to $N_\mathrm{upper}$, corresponding to $r_N$ varying from \SI{3}{nm} to \SI{8}{nm}. In both \textbf{(b)}~two and \textbf{(c)}~three dimensions, $r_\mathrm{deloc}$, estimated using \cref{eqn:IPR-rdeloc_conversion} and \cref{eqn:Effective_IPR}, decays as a function of $N$. The vertical lines correspond to $N_\mathrm{lower}$ (brown), $N_\mathrm{estim}$ (red), and $N_\mathrm{upper}$ (yellow), and the corresponding $r_\mathrm{deloc}$ is read off where these lines intersect the appropriate curve. Results are calculated using randomly initialised CT states, with $\sigma=\SI{150}{meV}$, $\lambda=\SI{200}{meV}$, and $T=\SI{300}{K}$.}
    \label{fig:Neighbourhood_rdelocs}
\end{figure}

Transport from these initial states is then modelled on the lattice by kinetic Monte Carlo, which gives the probability and the time taken for each hop~\cite{Bassler1993}. In conventional KMC, the hopping rates are usually given by nearest-neighbour Marcus rates~\cite{Marcus1956},
\begin{equation}
    \label{eqn:Marcus_rate}
    k_{if}^\mathrm{Marcus}=\frac{2\pi}{\hbar} \frac{J^2}{\sqrt{4\pi\lambda k_\mathrm{B}T}}\exp\left(-\frac{\left(\Delta E_{if}+\lambda\right)^2}{4\lambda k_\mathrm{B}T}\right),
\end{equation}
where $J$ is the electronic coupling between neighbouring sites, $\lambda$ is the reorganisation energy, $T$ is the temperature, and the energy between the final and initial sites is
\begin{equation}
    \Delta E_{if}=E_f-E_i-\frac{e^2}{4\pi\epsilon_0\epsilon_rd_{if}},
\end{equation}
where $E_i$ and $E_f$ are the energies of the initial and final sites, $d_{if}$ is the distance between them, $e$ is the electron charge, $\epsilon_0$ is the permittivity of free space, and $\epsilon_r$ is the dielectric constant (we take $\epsilon_r=3.5$).

By contrast, in jKMC, the hopping rate is calculated by assuming that centred on each lattice site is a spherically delocalised polaron of radius $r_\mathrm{deloc}$, described by \cref{eqn:Spherical_polaron_approximation}~\cite{Willson2023}. Applying this approximation in the secular polaron-transformed Redfield equation~\cite{Lee2015} leads to a correction of the Marcus rate,
\begin{equation}
    \label{eqn:jKMC_rate_Marcus}
    k_{\nu\nu'}^{\mathrm{jKMC}}=k_{\nu\nu'}^\mathrm{Marcus}\xi_{\nu\nu'},
\end{equation}
where $k_{\nu\nu'}^\mathrm{Marcus}$ is the Marcus rate of \cref{eqn:Marcus_rate} from polaron $\nu$ to $\nu'$ as if they were nearest neighbours and $\xi_{\nu\nu'}$ is a distance-dependent delocalisation correction,
\begin{equation}
    \label{eqn:Delocalisation_correction}
    \xi_{\nu\nu'}= A^4\sum_{\langle i,j\rangle} \exp\left(-\frac{2(d_{i\nu }+d_{j\nu'})}{r_\mathrm{deloc}}\right),
\end{equation}
where the sum runs over nearest-neighbour pairs of sites $i$ and $j$~\cite{Willson2023}. The delocalisation correction can be simplified in the limit of small delocalisation ($r_\mathrm{deloc}\ll a$) to yield simplified jKMC,
\begin{equation}
    \xi_{\nu\nu'}^\mathrm{simplified}=\frac{d_{\nu\nu'}}{a}\exp\left(-\frac{2\left(d_{\nu\nu'}-a\right)}{r_\mathrm{deloc}}\right),
\end{equation}
where $d_{\nu\nu'}$ is the hopping distance and $a$ is the lattice parameter, which we assume to be \SI{1}{nm}. Simplified jKMC is therefore similar to inserting the commonly used phenomenological delocalisation correction $e^{-2\gamma d_{if}}$ into the Marcus rate, and can be thought of as a rigorous justification of this correction in the limit of small delocalisation.

The competing process to separation is recombination, whose rates are also affected by delocalisation. Conventional KMC models allow recombination to occur at a rate $R_\mathrm{rec}$ (which we set to \SI{e10}{s^{-1}}) when the charges form a CT state at the interface~\cite{Groves2013,Kaiser2018}. Delocalisation facilitates long-range recombination of the electron and hole polarons beyond nearest neighbours in a CT state. The delocalised recombination rate in jKMC is derived from Fermi's golden rule following previous work~\cite{Tempelaar2016,Taylor2018,Balzer2022}, where using the spherical-polaron approximation gives
\begin{equation}
    \label{eqn:Recombination}
    k_{\nu_D\nu_A}^\mathrm{rec}=R_\mathrm{rec}A^4\left(\sum_{\substack{i,j\in\mathrm{CT} \\ i\in D, j\in A }}\exp\left(-\frac{d_{\nu_Di}+d_{\nu_Aj}}{r_\mathrm{deloc}}\right)\right)^2, 
\end{equation}
where $\nu_D$ is a hole polaron in the donor, $\nu_A$ is an electron polaron in the acceptor, and the sum runs over interfacial CT states consisting of donor site $i$ and acceptor site $j$ (full derivation in the Supporting Information). Simplifying this rate in the limit of small delocalisation ($r_\mathrm{deloc}\ll a$) for use in simplified jKMC yields
\begin{equation}
    \label{eqn:Recombination_simplified}
    k_{\nu_D\nu_A}^\mathrm{rec, simplified}=R_\mathrm{rec}\exp\left(-\frac{2(d_{\nu_D\nu_A}-a)}{r_\mathrm{deloc}}\right),
\end{equation}
where $d_{\nu_D\nu_A}$ is the distance between the donor and acceptor polarons (see Supporting Information for details).

The delocalisation radius $r_\mathrm{deloc}$ that parametrises jKMC hopping and recombination rates must be chosen to capture the effective polaron delocalisation during the separation process. We quantify the delocalisation using the inverse participation ratio (IPR), which, for a spherical state, is 
\begin{equation}
    \label{eqn:IPR-rdeloc_conversion}
    \mathrm{IPR}_\mathrm{jKMC}=A^{-4}\Bigg(\sum_i \exp \left(-\frac{4d_{i\nu}}{r_\mathrm{deloc}}\Bigg)\right)^{-1}.
\end{equation}
To choose $r_\mathrm{deloc}$ so that it is representative of the delocalisation in the system, we set $\mathrm{IPR}_\mathrm{jKMC}$ to be equal to an estimate of the thermally weighted polaron IPR during charge separation in dKMC. The latter is calculated as a Boltzmann expectation value of the IPR over a number $N$ of polaron eigenstates,
\begin{equation}
    \mathrm{IPR}_\mathrm{jKMC}(N)=\left\langle\frac{1}{Z} \sum_{\nu=1}^N \mathrm{IPR}_\nu \exp\left(-\frac{E_\nu}{k_\mathrm{B}T}\right)\right\rangle,
    \label{eqn:Effective_IPR}
\end{equation}
where $\mathrm{IPR}_\nu$ and $E_\nu$ are the polaron IPRs and energies obtained from the diagonalisation of the dKMC single-particle Hamiltonian~\cite{Balzer2021,Willson2023}, $Z=\sum_{\nu=1}^N \exp\left(-E_\nu/k_\mathrm{B}T\right)$ is the partition function, and the average $\langle \cdot \rangle$ is taken over an ensemble of disordered energetic landscapes (1000 in our calculations)~\cite{Willson2023}. The thermal average is taken over a finite number $N$ of polarons since the charge does not fully sample the disordered DOS during the charge separation process.

\begin{figure*}
    \centering
    \includegraphics[width=\textwidth]{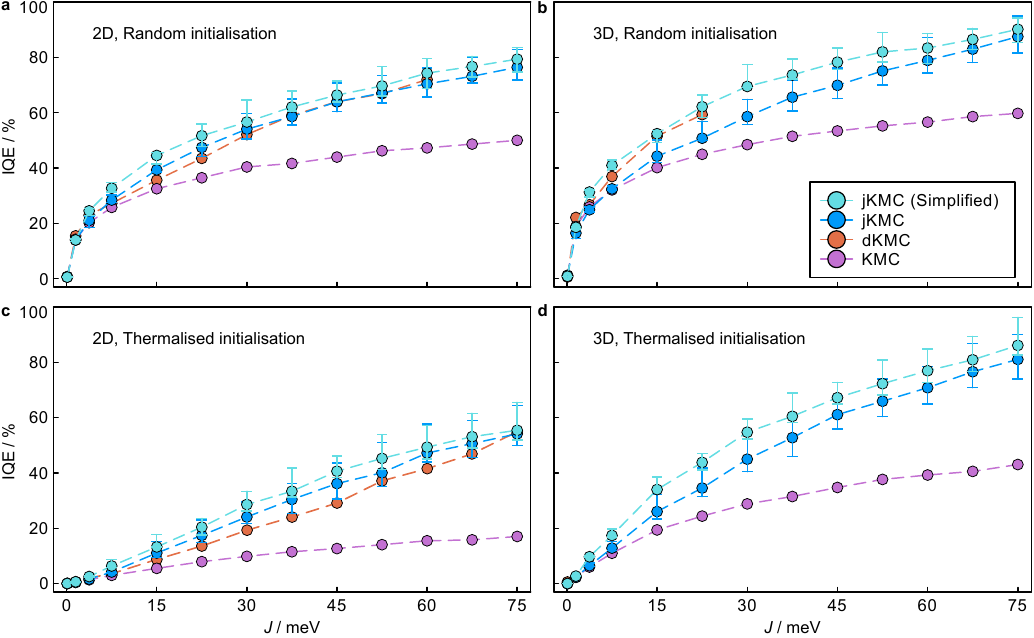}
    \caption{\textbf{jKMC reproduces the significant delocalisation enhancements of IQE seen in dKMC}, in both \textbf{(a, c)}~two and \textbf{(b, d)}~three dimensions. Increasing the coupling $J$ produces greater delocalisation, leading to much higher IQEs, as shown by dKMC, jKMC and simplified jKMC. jKMC reproduces most dKMC IQEs within the range due to uncertainty in $r_\mathrm{deloc}$ (error bars, defined in \cref{fig:Neighbourhood_rdelocs}). jKMC is robust to the initial condition, agreeing with dKMC for charges initialised in interfacial CT states selected both randomly (a,~b) and with thermalised weightings (c,~d). dKMC is too computationally expensive in three dimensions to model the full range of $J$ shown. Simplified jKMC slightly overestimates both jKMC and dKMC, but remains significantly more accurate than conventional KMC. Results are calculated for $\sigma=\SI{150}{meV}$, $\lambda=\SI{200}{meV}$, and $T=\SI{300}{K}$.}
    \label{fig:IQE_results}
\end{figure*}

We estimate $N$ based on a typical number of sites accessible to either of the charges during the charge separation (\cref{fig:Neighbourhood_rdelocs}a). We consider the charges to be free if their separation exceeds a fixed distance $r_\mathrm{sep}=\SI{5}{nm}$~\cite{Balzer2022}, so we set $N$ to be equal to the number of sites inside a hemisphere with radius $r_N=r_\mathrm{sep}=\SI{5}{nm}$, i.e., $N=\lfloor \pi (r_N/a)^2/2 \rfloor$ in 2D and $N=\lfloor 4\pi (r_N/a)^3/6 \rfloor$ in 3D. For $r_N=\SI{5}{nm}$, we obtain the central estimates $N_\mathrm{estim}=39$ in two dimensions and $N_\mathrm{estim}=262$ in three dimensions. However, because these are rough estimates, we account for uncertainty in the appropriate choice of $N$ by using lower and upper estimates of $N$ to construct a range of plausible $r_\mathrm{deloc}$ for our results. As shown in \cref{fig:Neighbourhood_rdelocs}a, we choose $N_\mathrm{lower}$ and $N_\mathrm{upper}$ based on  $r_N=\SI{3}{nm}$ and $r_N=\SI{8}{nm}$, respectively. Using this range of $N$, we calculate a range of IPRs using \cref{eqn:Effective_IPR}, then solve \cref{eqn:IPR-rdeloc_conversion} for the range of $r_\mathrm{deloc}$ that give the same range of IPRs (\cref{fig:Neighbourhood_rdelocs}b--c). We use this range of $r_\mathrm{deloc}$ for both randomly and thermally initialised CT states. As we discuss below, the results obtained within this range of $r_\mathrm{deloc}$ are similar, meaning that the uncertainty in $N$ or $r_\mathrm{deloc}$ does not affect our qualitative conclusions.

The simulation can end in two ways. The charges are considered free if they become separated by at least $r_\mathrm{sep}=\SI{5}{nm}$ and the percentage of simulations that end this way is the IQE. Alternatively, charges can recombine across the donor-acceptor interface. In numerical simulations, we also impose a maximum hopping limit to avoid infinite loops, and if the limit is exceeded, the charges are also considered to have recombined. We set the cutoff at \num{10000} hops, which is sufficient to converge the IQEs to significantly less than other sources of error.

\section{Results \& Discussion}

jKMC reproduces the significant enhancements in charge separation efficiency due to delocalisation seen in dKMC (\cref{fig:IQE_results}). For randomly initialised CT states in both two and three dimensions, jKMC predicts IQEs that are as much as double the corresponding KMC values, for our chosen parameters. For thermally initialised CT states in two dimensions with the same parameters, jKMC predicts a fivefold increase in IQE due to delocalisation. Therefore, jKMC captures the significant trends first predicted by dKMC for both initial conditions.

jKMC is reasonably accurate quantitatively, striking a balance between accuracy and speed. For both initialisation conditions, jKMC slightly overestimates the majority of 2D dKMC IQEs and slightly underestimates the 3D dKMC IQEs (where those can be calculated). However, most dKMC IQEs fall within the uncertainty ranges of jKMC. The uncertainty ranges are sufficiently narrow to offer a significant separation from KMC predictions, and even taking the minimum amount of estimated delocalisation leads to significantly higher IQEs than in KMC. The uncertainty range is wider in three dimensions, due to the larger range of $r_\mathrm{deloc}$ (shown in \cref{fig:Neighbourhood_rdelocs}c). 

We obtain good results in two dimensions using the same set of $r_\mathrm{deloc}$ for thermally and randomly initialised CT states (\cref{fig:IQE_results}a,c), despite the typical initial delocalisation of thermalised states being smaller than that of randomly initialised ones. This insensitivity to the delocalisation of the initial state likely occurs because the separating charges must travel through many other states before separating, suggesting that the overall IQE is determined by the effective delocalisation along their entire trajectory, which, to a good approximation, is captured by the same $r_\mathrm{deloc}$ regardless of initial differences. The insensitivity to the initial delocalisation further justifies our assumption that the initial CT state is separable, an assumption that neglects the additional localisation caused by the Coulomb attraction between the charges.

The low computational cost of jKMC allows its use for calculations impossible in dKMC. In particular, dKMC is too computationally expensive in three dimensions for moderately delocalised states or to thermally initialise over a significant number of CT states~\cite{Balzer2022}. However, jKMC can describe both regimes easily, where it continues to predict high delocalisation enhancements (\cref{fig:IQE_results}b,d).

Despite its simplicity, simplified jKMC also provides reasonably accurate IQEs, which are still significantly more accurate than the corresponding KMC IQEs (\cref{fig:IQE_results}). The uncertainty in the simplified jKMC IQEs due to uncertainty in $r_\mathrm{deloc}$ is smaller than that of the jKMC IQEs, indicating that simplified jKMC is more stable to variations of $r_\mathrm{deloc}$. This stability occurs because the hopping and recombination rates in simplified jKMC are scaled by the same exponential factor ($e^{-2\left(d_{\nu\nu'}-a\right)/r_\mathrm{deloc}}$), and thus separation and recombination rates increase or decrease identically when varying $r_\mathrm{deloc}$. These competing effects cancel out in the ratio defining the IQE to a higher degree in simplified jKMC than in jKMC.

\begin{figure*}
    \centering
    \includegraphics[width=\textwidth]{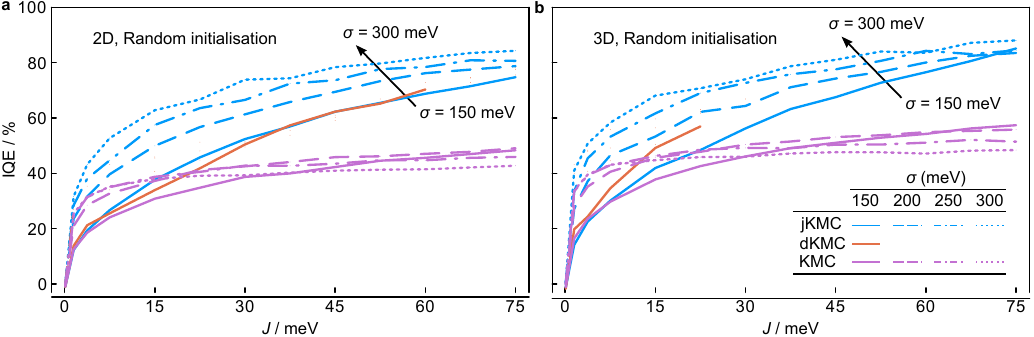}
    \caption{\textbf{The low computational cost of jKMC allows extensive parameter scans of organic semiconductors.} In both \textbf{(a)}~two and \textbf{(b)}~three dimensions, jKMC is more accurate than KMC and significantly cheaper than dKMC, with jKMC able to model the full parameter range shown. Results are for $N=N_\mathrm{estim}$, $\lambda=\SI{200}{meV}$, and $T=\SI{300}{K}$.}
    \label{fig:Parameter_scan_results}
\end{figure*}

The low computational cost of jKMC allows it to perform extensive parameter scans that may assist in computational screening. \Cref{fig:Parameter_scan_results} compares IQEs for jKMC, dKMC, and KMC for systems between moderate (\SI{150}{meV}) and high (\SI{300}{meV}) disorders. jKMC predicts that higher disorders lead to larger IQEs, as opposed to KMC, which shows the opposite trend at high couplings.

The computational cost of jKMC depends on how it is performed. If $r_\mathrm{deloc}$ is treated as a fitting parameter (like $\gamma$), then jKMC has the same cost as conventional KMC models with non-nearest-neighbour hopping. If $r_\mathrm{deloc}$ is calculated from the material parameters, then jKMC is slightly more expensive than KMC because some diagonalisations are used to calculate $r_\mathrm{deloc}$. However, the diagonalisations required to calculate $r_\mathrm{deloc}$ for jKMC are much simpler and fewer than those needed in dKMC. First, we diagonalise only single-particle Hamiltonians instead of the much larger two-particle Hamiltonians, which allows us to use the same values of $r_\mathrm{deloc}$ that were used previously for single-particle charge transport~\cite{Willson2023}. Second, for each energy landscape, these diagonalisations only need to be carried out once, as opposed to having to be repeated for every timestep in dKMC. 

Extensions that could easily be added to jKMC in the future would enable the exploration of even wider parameter spaces for materials discovery. For example, jKMC could be easily extended to non-identical parameters for the donor and acceptor. The assumed planar heterojunction could also be relaxed to study other morphologies. Modifications to the interface layers~\cite{deSousa2020} could be performed to study the effects of interfacial dipoles in driving charge separation~\cite{Yi2009,Yang2014,Ran2017,Khan2021}, while including important effects from delocalisation. Including more than two particles would enable modelling charge-density effects and non-geminate recombination under an accurate mesoscopic treatment of delocalisation.

Future developments to jKMC could also relax the assumptions or approximations used in deriving the jKMC rate~\cite{Willson2023}, leading to a more widely applicable method. For example, materials with anisotropic electronic couplings could be modelled by replacing the assumption of identical spherical polarons with identical ellipsoidal polarons. Similarly, jKMC could be extended to include a treatment of long-range electronic couplings, non-Gaussian DOS, and spatial site-energy correlations.

\section{Conclusion}
We demonstrated that jKMC can reproduce the large enhancements to charge-separation efficiencies caused by delocalisation, as first seen in dKMC. Furthermore, the low computational cost of jKMC has allowed us to carry out delocalised charge-separation calculations and parameter scans in regimes beyond the limits of dKMC. In three dimensions, these calculations are able to explain near-unity IQEs even in materials with significant disorder. 
The functional form of jKMC is a simple modification of the Marcus rate, which can be used to incorporate delocalisation into any charge-separation KMC model. In the future, we anticipate using the mesoscopic properties calculated using jKMC---mobilities, recombination coefficients, and IQEs---as input parameters for drift-diffusion simulations, thereby unraveling device-level effects of delocalisation.

\begin{acknowledgments}
We were supported by the Australian Research Council (DP220103584), by a Westpac Scholars Trust Future Leaders Scholarship, by an Australian Government Research Training Program scholarship, and by computational resources from the Sydney Informatics Hub (Artemis) and the Australian Government's National Computational Infrastructure (Gadi) through the National Computational Merit Allocation Scheme.
\end{acknowledgments}

\bibliographystyle{achemso}
\bibliography{bib}

\setcounter{subsection}{0}
\renewcommand{\thesubsection}{S\arabic{subsection}}%
\setcounter{equation}{0}
\renewcommand{\theequation}{S\arabic{equation}}%
\setcounter{figure}{0}
\renewcommand{\thefigure}{S\arabic{figure}}%

\section*{Supporting Information}

\subsection{Derivation of the jKMC recombination rate}
\label{SI:Recombination}

The recombination rate in jKMC is obtained by simplifying a Fermi-golden-rule expression using the same approximations as those made in deriving the jKMC hopping rate. Following~\cite{Balzer2022}, we start with the recombination rate between the joint polaron-pair state $\ket{\nu}$ and the ground state $\ket{g}$, given by Fermi's golden rule as
\begin{equation}
    \label{eqn:FGR}
    k^\mathrm{rec}_{\nu}=2\pi\left|\bra{\nu}H\ket{g}\right|^2\rho_\mathrm{rec},
\end{equation}
where $H$ is the Hamiltonian connecting the polaron-pair to the ground state, and $\rho_\mathrm{rec}$ is the density of states. Inserting into \cref{eqn:FGR} a resolution of the identity over the donor and acceptor sites $\ket{i,j}$ gives
\begin{equation}
    k^\mathrm{rec}_{\nu}=2\pi\left|\sum_{\substack{i,j\in\mathrm{CT} \\ i\in D, j\in A }}\braket{{\nu}|{i,j}}\bra{i,j}H\ket{g}\right|^2\rho_\mathrm{rec},
    \label{eqn:FGR_recombination}
\end{equation}
where the sum only runs over pairs of interfacial CT sites $\ket{i,j}$ because we assume that $\bra{i,j}H\ket{g}=0$ otherwise, i.e., that there is no recombination except from nearest-neighbour sites. Assuming that each CT state is equally coupled to the ground state, $\bra{i,j}H\ket{g}=J_\mathrm{rec}$, we obtain~\cite{Balzer2022}
\begin{equation}
    k^\mathrm{rec}_{\nu}=R_\mathrm{rec}\left|\sum_{\substack{i,j\in\mathrm{CT} \\ i\in D, j\in A }}\braket{{\nu}|{i,j}}\right|^2,
    \label{eqn:FGR_recombination_const_coupling}
\end{equation}
where $R_\mathrm{rec}=2\pi J_\mathrm{rec}^2\rho_\mathrm{rec}$ is the recombination rate for a localised CT state.
Therefore, \cref{eqn:FGR_recombination_const_coupling} gives the delocalised recombination rate as the localised recombination rate $R_\mathrm{rec}$ multiplied by a delocalisation correction. 

The delocalisation correction can be evaluated for the spherical polarons used in jKMC by assuming that the joint polaron-pair state can be separated into independent electron and hole states ($\ket{\nu}=\ket{\nu_D,\nu_A}$, where $\nu_D$ is the hole polaron in the donor and $\nu_A$ is the electron polaron in the acceptor), to which we apply the spherical polaron approximation introduced in the main text,
\begin{equation}
    \braket{{\nu_D,\nu_A}|{i,j}}=A^2\exp\left(-\frac{d_{\nu_Di}}{r_\mathrm{deloc}}\right)\exp\left(-\frac{d_{\nu_Aj}}{r_\mathrm{deloc}}\right),
    \label{eqn:Spherical_polaron_approximation_CT}
\end{equation}
where $A=\left(\sum_i \exp\left(-2d_{i \nu}/r_\mathrm{deloc}\right)\right)^{-1/2}$ is the normalisation constant, $d_{\nu_Di}$ is the distance between the centre of polaron $\nu_D$ and site $i$ (and likewise for $d_{\nu_Aj}$), and $r_\mathrm{deloc}$ is the delocalisation radius. Substituting \cref{eqn:Spherical_polaron_approximation_CT} into \cref{eqn:FGR_recombination_const_coupling} yields the jKMC recombination rate,
\begin{equation}
    k_{\nu_D\nu_A}^\mathrm{rec}=R_\mathrm{rec}A^4\left(\sum_{\substack{i,j\in\mathrm{CT} \\ i\in D, j\in A }}\exp\left(-\frac{d_{\nu_Di}+d_{\nu_Aj}}{r_\mathrm{deloc}}\right)\right)^2.
    \label{eqn:Recombination_jKMC}
\end{equation}

\begin{figure}[b]
    \centering
    \includegraphics[width=0.85\columnwidth]{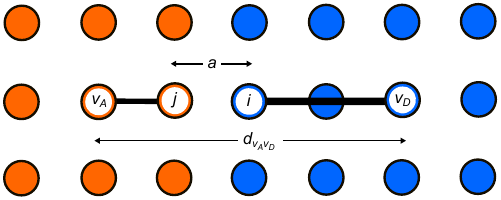}
    \caption{\textbf{The arrangement of sites whose contribution dominates the jKMC recombination rate.} The dominant terms in \cref{eqn:Recombination_jKMC} are those with the smallest distance $d_{\nu_Di}+d_{\nu_Aj}$ (shown in black). On the cubic lattice, the simplest case to consider is if the donor and acceptor polarons $\nu_D$ and $\nu_A$ lie in the same row of the lattice, as shown above. In that case, the distance is minimised and equals $d_{\nu_D\nu_A}-a$ when the nearest neighbours $i$ and $j$ are collinear with $\nu_D$ and $\nu_A$. 
    We use the same value of $d_{\nu_D\nu_A}-a$ even if $\nu_D$ and $\nu_A$ do not lie in the same row, an approximation that is justified because the particular shape of the lattice is not an essential feature of the jKMC model.
    }
    \label{fig:Simplified_jKMC_CT}
\end{figure}

This rate may be further simplified in the limit of small delocalisation ($r_\mathrm{deloc}\ll a$) for use with simplified jKMC. For small $r_\mathrm{deloc}$, the sum in \cref{eqn:Recombination_jKMC} is dominated by terms that minimise the distance $d_{\nu_Di}+d_{\nu_Aj}$, where $i$ and $j$ are sites of interfacial CT states. As shown in \cref{fig:Simplified_jKMC_CT}, there is one dominant term, with the minimal $d_{\nu_Di}+d_{\nu_Aj}=d_{\nu_D\nu_A}-a$. Keeping only this term, and noting that $A=1$ in the limit of localised charges, we obtain the simplified jKMC recombination rate,
\begin{equation}
    k_{\nu_D\nu_A}^\mathrm{rec, simplified}=R_\mathrm{rec}\exp\left(-\frac{2(d_{\nu_D\nu_A}-a)}{r_\mathrm{deloc}}\right).
\end{equation}

\end{document}